\documentclass[a4paper,11pt]{article}
\usepackage{latexsym,epsfig}
\usepackage{amsmath,amssymb,amsthm}
\usepackage{graphicx}
\usepackage{color}
\usepackage{subfig}
\usepackage{listings}
\usepackage{courier}
\usepackage{verbatim}
\usepackage[margin=1.1in]{geometry}
\usepackage{tabularx}
\usepackage{epstopdf}
\usepackage{url}
\usepackage{hyperref}

\newtheorem{thm}{Theorem}[section]
\newtheorem{prop}[thm]{Proposition}
\newtheorem{lem}[thm]{Lemma}
\newtheorem{cor}[thm]{Corollary}

\newtheorem{obs}[thm]{Observation}

\newtheorem{problem}{Problem}

\newenvironment{proofof}[1]
{\medskip\noindent{\em Proof of #1:}\hspace{1mm}}
{\hfill$\Box$\medskip}
\def\L{{\cal L}}

\begin{document}
	
	\title{
		The Zarankiewicz Problem for Polygon Visibility Graphs
	}
	
	\author{
		Eyal Ackerman\thanks{Department of Mathematics, Physics and Computer Science,
			University of Haifa at Oranim, 	Tivon 36006, Israel. \texttt{ackerman@math.haifa.ac.il}}\and
		Bal\'azs Keszegh\thanks{HUN-REN Alfréd Rényi Institute of Mathematics and ELTE Eötvös Loránd University, Budapest, Hungary. 
			Research supported by the National Research, Development and Innovation Office -- NKFIH under the grant K 132696 and by the ERC Advanced Grant ``ERMiD''. This research has been implemented with the support provided by the Ministry of Innovation and Technology of Hungary from the National Research, Development and Innovation Fund, financed under the  ELTE TKP 2021-NKTA-62 funding scheme.}
	}

	\date{}
	\maketitle

	\begin{abstract}
		We prove a quasi-linear upper bound on the size of $K_{t,t}$-free polygon visibility graphs.
		For visibility graphs of star-shaped and monotone polygons we show a linear bound.
		In the more general setting of $n$ points on a simple closed curve and visibility pseudo-segments, we provide an $O(n \log n)$ upper bound and an $\Omega(n\alpha(n))$ lower bound.
	\end{abstract}

	\section{Introduction}
	\label{sec:intro}
	
	Determining the maximum number of edges in a bipartite graph which does not contain the complete bipartite graph $K_{s,t}$ as a subgraph is known as the Zarankiewicz Problem and is a central problem in Extremal Graph Theory.
	Since every graph has a bipartite subgraph with at least half of the edges, it is essentially equivalent to consider this problem for arbitrary host graphs.
	It is also customary to study the symmetric case, that is, the maximum number of edges in a $K_{t,t}$-free $n$-vertex graph.
	The classical K\H{o}v\'ari-S\'os-Tur\'an Theorem~\cite{Kvri1954OnAP} gives the upper bound $O_t(n^{2-1/t})$ which is known to be asymptotically tight in some cases, namely, for $t=2,3$.
	It is a major open problem in combinatorics to determine whether this bound is tight for greater values of $t$.
	
	The K\H{o}v\'ari-S\'os-Tur\'an upper bound can be improved for certain classes of host graphs.
	For example, Fox, Pach, Sheffer, Suk and Zahl~\cite{FoxPSSZ} provided an $O_{t,d}(n^{2-1/d})$ upper bound for $K_{t,t}$-free graphs of VC-dimension at most $d$ (see also~\cite{FranklK21,JanzerP24,KellerS2024}), whereas Gir\~ao and Hunter~\cite{GH24} and Bourneuf, Buci\'c, Cook, and Davies~\cite{BBCD24} proved an $O_t(n^{2-\varepsilon(H)})$ bound for $K_{t,t}$-free graphs that avoid a given bipartite graph $H$ as an induced subgraph. 
	For other results concerning graphs of certain structural properties see~\cite{HunterMST25} and the references within and also the recent survey by Du and McCarty~\cite{degree-boundeness}.
	
	A different yet related line of research studies $K_{t,t}$-free graphs that arise from geometry, usually incidence or intersection graphs of geometric objects.
	For example, the incidence graph between $n$ points and $n$ lines\footnote{In this graph the vertices correspond to points and lines and there is an edge between a vertex that represents a point and a vertex that represents a line if that point lies on that line.} is $K_{2,2}$-free and hence has $O(n^{3/2})$ edges by the K\H{o}v\'ari-S\'os-Tur\'an Theorem.
	The celebrated Szemer\'edi-Trotter Theorem~\cite{Szemerdi1983ExtremalPI} improves this bound to $O(n^{4/3})$ which is tight.
	Moreover, in some geometric settings linear or almost linear upper bounds are known, for instance, the maximum size of a $K_{t,t}$-free intersection graph of $n$ curves in the plane\footnote{In this graph, known as \emph{string graph}, every curve is represented by a vertex and every pair of intersecting curves by an edge.} is $O_t(n)$~\cite{FP14}.
	For further results on the Zarankiewicz Problem in geometric settings see the recent survey by Smorodinsky~\cite{smorodinsky2024}.
	
	In this work we also study the Zarankiewicz Problem in a geometric setting, however,
	rather than incidence or intersection graphs we consider \emph{visibility} graphs, and more specifically,
	\emph{polygon visibility} graphs, their generalizations and some special cases.
	Given a simple polygon $P$ and two points $p,q\in P$ we say that $p$ and $q$ are \emph{mutually visible} with respect to $P$ (or that they `see' each other), if the straight-line segment $\overline{pq}$ is disjoint from the exterior of $P$.	
	The \emph{visibility graph} of $P$ consists of vertices that correspond to the vertices of $P$ and edges that correspond to pairs of mutually visible vertices.
	We prove the following quasi-linear bound on the size of a $K_{t,t}$-free visibility graph of a simple polygon.

\begin{thm}\label{thm:polygon-ub}
	For every integer $t>0$ there is a constant $c_t$ such that the following holds.
	Let $P$ be a simple $n$-gon and let $G$ be the visibility graph of $P$.
	If $G$ is $K_{t,t}$-free, then it has $O(n2^{\alpha(n)^{c_t}})$ edges, 
	where $\alpha(n)$ is the inverse Ackermann function.
\end{thm}
	
	The previous best bound we are aware of follows from a result of Agarwal, Alon, Aronov and Suri~\cite{Agarwal1994}, stating that a polygon visibility graph always has a clique cover\footnote{A \emph{clique cover} of a graph consists of clique and biclique subgraphs whose union contains every edge of the graph. The size of the cover is sum of the number of vertices over all the (bi)cliques.} of size $O(n\log^3 n)$. Combining this with an observation of Do~\cite{Do2019} by which the number of edges in a $K_{t,t}$-free bipartite graph is at most $2t$ times the size of its biclique cover,\footnote{See the discussion following~\cite[Corollary~1]{Do2019}} we get an $O(tn\log^3 n)$ upper bound.
	We note, however, that Agarwal et al.~\cite{Agarwal1994} also show an $\Omega(n\log n)$ worst case lower bound on the size of the clique cover of a simple $n$-gon, implying that further arguments are needed for using this method to get a better bound than the one in Theorem~\ref{thm:polygon-ub}.
		
	\medskip
	A simple polygon is \emph{star-shaped} if it contains a point that sees every other point in the polygon.
	If every vertical line intersects the boundary of a polygon at most twice, then this polygon is \emph{$x$-monotone.}
	We obtain linear upper bounds on the size of $K_{t,t}$-free visibility graphs of star-shaped polygons and of $x$-monotone polygons.
	
\begin{thm}\label{thm:star-shaped}
	Let $t>0$ be an integer, let $P$ be a star-shaped polygon and let $V$ be a set of $n$ points on its boundary.
	If the visibility graph of $V$ with respect to $P$ is $K_{t,t}$-free, then it contains $O_t(n)$ edges.
\end{thm}
	
\begin{thm}\label{thm:x-monotone}
	Let $t >0$ be an integer, let $P$ be an $x$-monotone polygon and let $V$ be a set of $n$ points on its boundary.
	If the visibility graph of $V$ with respect to $P$ is $K_{t,t}$-free, then it contains $O_t(n)$ edges.
\end{thm}
	 
	Note that the bounds for star-shaped and $x$-monotone polygons hold for any set of $n$ points on their boundaries, whereas in Theorem~\ref{thm:polygon-ub} every vertex of the polygon is a vertex of the visibility graph.
	It is known that polygon visibility graphs are not hereditary, that is, an induced subgraph of a polygon visibility graph may not be the visibility graph of some other polygon.
	Induced subgraphs of polygon visibility graphs are equivalent to \emph{curve visibility graphs} which are visibility graphs of points on a Jordan curve.\footnote{Two vertices in such a graph are adjacent if the straight-line segment between their corresponding points is disjoint from the exterior of the curve.} 
	Du and McCarty~\cite{degree-boundeness} mention the following open problem which was the motivation for this work.

\begin{problem}[\cite{degree-boundeness}]\label{prob:curve}
	Does every $n$-vertex $K_{t,t}$-free curve visibility graph have $O_t(n)$ edges?
\end{problem}

	Although we do not settle Problem~\ref{prob:curve}, we were able to provide a negative answer for the more general class of \emph{curve pseudo-visibility graphs} which are defined as follows (see~\cite{Coloring23}).
	Let $K$ be a Jordan curve, let $V$ be a set of points on $K$ and let $\L$ be a set of pseudolines such that every pair of points in $V$ lies on a pseudoline from $\L$.
	The \emph{curve pseudo-visibility graph}  $G_\L(K,V)$ consists of the vertex set $V$ and edges between every pair of points such that the segment of the pseudoline connecting them is disjoint from the exterior of $K$.

\begin{thm}\label{thm:pseudo-visibility-lb}
	For every positive integer $n$ there is a set of $n$ points $V$, a Jordan curve $K$ and a set of pseudolines $\L$ such that the curve pseudo-visibility graph $G_{\L}(K,V)$ is $K_{3,3}$-free and has $\Omega(n\alpha(n))$ edges, where $\alpha(n)$ is the inverse Ackermann function. 
\end{thm}

	Following a result of McCarty~\cite{McCarty21}, we may also conclude that the maximum number of edges in a $K_{2,2}$-free curve pseudo-visibility graph is also super-linear.
	\begin{cor}\label{cor:K_2,2}
		For every constant $c_2$ there is an $n$-vertex $K_{2,2}$-free curve pseudo-visibility graph with more than $c_2n$ edges.
	\end{cor}

	As for an upper bound, we have:

\begin{thm}\label{thm:pseudo-visibility-ub}
	Let $t>0$ be an integer and let $G_{\cal L}(K,V)$ be a $K_{t,t}$-free curve pseudo-visibility graph of a set of $n$ points $V$ that lie on a Jordan curve $K$ such that every pair of points in $V$ also lie on a pseudoline from a set of pseudolines $\L$.
	Then $G_{\cal L}(K,V)$ has $O_t(n\log n)$ edges.
\end{thm}

	\paragraph{Related work.}
	Visibility graphs of polygons (and other geometric creatures) have been studied extensively both from a combinatorial and a computational point of view, see, e.g.,~\cite{Ghosh97,GHOSH2007,GhoshG13,Hershberger89,CAGIRICI2024}.
	It is worth mentioning a recent result of Davies, Krawczyk, McCarty and Walczak~\cite{Coloring23} who proved that curve pseudo-visibility graphs are \emph{$\chi$-bounded}, that is, their chromatic number can be bounded from above in terms of their clique number.
	
	If the maximum size of any $K_{t,t}$-free $n$-vertex graph in a certain family of graphs is at most $c_tn$, where $c_t$ depends only on $t$, then this family of graphs is called \emph{degree-bounded}.\footnote{Since every $K_{t,t}$-free graph in the family has a vertex of a small degree -- at most $2c_t$.} 
	By a remarkable result of Gir\~ao and Hunter~\cite{GH24}, if a hereditary family of graphs is degree-bounded, then $c_t$ can be bounded by a polynomial function.
	For a further discussion on degree-boundedness see the recent survey of Du and McCarty~\cite{degree-boundeness}.

	\paragraph{Organization.} The bounds for curve pseudo-visibility graphs are proved in Section~\ref{sec:pseudo-visibility}.
	In Section~\ref{sec:polygon} we prove the bounds for polygon visibility graphs including the special cases of star-shaped and $x$-monotone polygons.

	\section{Terminology and tools}
	\label{sec:tools}
	
	An \emph{ordered graph} is a graph whose vertices are linearly ordered.
	In a \emph{cyclically-ordered graph} $G=(V,E,<)$ the vertices are ordered cyclically.
	For any vertex $v$ in $G$ we denote by $G_v=(V,E,<_v)$ the ordered graph such that $<_v$ is the linear order induced by $<$ when setting $v$ as the smallest element.
	
	Let $G=(V,E,<)$ be an ordered graph and let $A,B \subseteq V$ be two vertex subsets. 
	We write $A < B$ if $a < b$ for every $a \in A$ and $b \in B$.
	If $G$ is cyclically-ordered, then $A < B$ means that $A <_x B$ for some $x \in V$ (which also implies that $B < A$).
	In both cases $v < A$ stands for $\{v\} < A$.
	Two edges $(a,b),(c,d)$ in an (cyclically-) ordered graph \emph{cross} if $a<c<b<d$.\footnote{When we write $v_1 < v_2 < \ldots < v_k$ for some $k \ge 3$ vertices of a cyclically-ordered graph $G=(V,E,<)$, we mean that there is a vertex $x\in V$ such that $v_1 <_x v_2 <_x \ldots <_x v_k$.}
	The set of neighbors of a vertex $v$ in $G$ is denoted by $N_G(v)$.
		
\begin{lem}\label{lem:left-right-vertices}
	Every ordered graph $G=(V,E,<)$ has an ordered bipartite subgraph $G'=(L \cup R,E',<)$ such that $v < N_{G'}(v)$ for every vertex $v \in L$; $v > N_{G'}(v)$ for every vertex $v \in R$ and $|E'| \ge |E|/4$.
\end{lem}

\begin{proof}
	For every vertex $v$ we toss a fair coin.
	If the result is `heads', then we place $v$ in $L$ and delete all the edges $(u,v)$ such that $u<v$.
	Otherwise, if the result is `tails' we place $v$ in $R$ and delete all the edges $(u,v)$ such that $u>v$.
	Clearly the resulting subgraph $G'$ is bipartite, $v < N_{G'}(v)$ for every vertex $v \in L$ and $v > N_{G'}(v)$ for every vertex $v \in R$.
	Since every edge of $G$ survives with probability $1/4$, the expected number of edges in $G'$ is $|E|/4$.
	Therefore, there is a series of coin tosses that yields at least that many edges.
\end{proof}

	For points lying on a Jordan curve $K$ we denote by $<_K$ their clockwise cyclic order along $K$.
	The clockwise sub-curve from a point $p \in K$ to a point $q \in K$ is denoted by $p \leadsto_K q$.
	Any curve pseudo-visibility graph $G_\L(K,V)$ can naturally be considered as a cyclically-ordered graph where $<_K$ is the cyclic order of its vertices.
	Furthermore, edges cross in the sense of cyclically-ordered graphs if and only if they cross in the geometric sense in these graphs.
	This fact is trivial for polygon and curve visibility graphs, and is not hard to prove for curve pseudo-visibility graphs, see~\cite[Lemma 2.3]{Coloring23}.

\medskip
	Let $u$ and $v$ be two vertices of an ordered graph $G$ such that $u<v$.
	Following~\cite{Coloring23} we say that a sequence of edges $e_1,e_2,\ldots,e_k$ is a \emph{crossing sequence from $u$ to $v$} if $u$ is the smaller endpoint of $e_1$, $v$ is the greater endpoint of $e_k$, and $e_i$ crosses $e_{i+1}$, for every $k=1,2,\ldots,k-1$.
	If $G$ is cyclically-ordered, then $e_1,e_2,\ldots,e_k$ is a crossing sequence from $u$ to $v$, if it is a crossing sequence from $u$ to $v$ in $G_u$.
	
\begin{figure}
	\centering
	\includegraphics[width= 7cm]{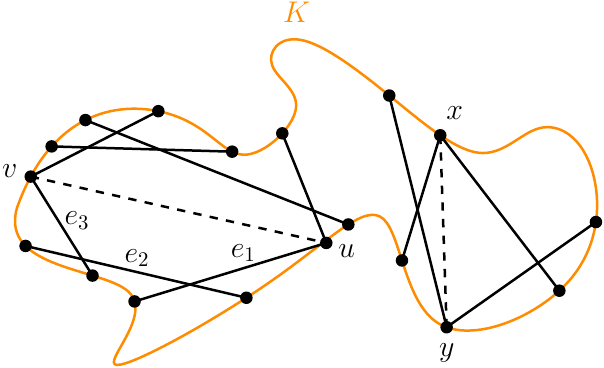}
	\caption{$e_1,e_2,e_3$ is a crossing sequence from $u$ to $v$ in this curve visibility graph. There is also a crossing sequence from $v$ to $u$ and therefore they must see each other. $x$ and $y$ form a double cherry.}
	\label{fig:crosing-sequence}
\end{figure}

	It is easy to see that if there are crossing sequences from $u$ to $v$ and from $v$ to $u$ in a curve visibility graph, then these vertices must be adjacent (see Figure~\ref{fig:crosing-sequence}).
	Proving this for curve pseudo-visibility graphs is more tricky.
	
\begin{lem}[{\cite[Proposition~3.2]{Coloring23}}]\label{lem:crossing-sequences}
	For every curve pseudo-visibility graph and every pair of vertices in it, if there are crossing sequences
	from one vertex to the other vertex and vice versa, then these vertices are adjacent in the graph.
\end{lem}

	A \emph{convex geometric graph} is a graph drawn such that its vertices are in convex position and its edges are straight-line segments.
	The following result of Capoyleas and Pach~\cite{CP92} was proved for convex geometric graphs, however, its proof applies verbatim for cyclically-ordered graphs.

\begin{thm}[\cite{CP92}]\label{thm:Capoyleas-Pach}
	Let $G$ be an $n$-vertex cyclically-ordered graph and let $k \le (n-1)/2$ be a positive integer.
	If $G$ does not contain $k+1$ pairwise crossing edges, then it has at most $2kn - {{2k+1}\choose{2}}$ edges.
\end{thm}

	An $n$-vertex ordered graph $G$ can be represented by an $n\times n$ $0$-$1$ matrix whose rows and columns correspond to the vertices of $G$ and are ordered according to their order and the $(i,j)$-entry is $1$ iff the $i$th and $j$th vertices of $G$ are adjacent, for every $1 \le i < j \le n$.
	A bipartite ordered graph $(A\cup B,E)$ can be partially represented by a $0$-$1$ matrix whose rows and columns correspond to $A$ and $B$, respectively. 
	Note that this representation does not capture the order over vertices of different classes, unless it is known that $A < B$.
	
	We say that a $0$-$1$ matrix $M$ \emph{contains} a $0$-$1$ matrix $P$, if $P$ can be obtained from $M$ by deleting some rows and columns and changing some $1$-entries to $0$.
	$M$ \emph{avoids} $P$ or is \emph{$P$-free} if it does not contain $P$.
	Bounding the maximum number of $1$-entries in $n \times n$ $0$-$1$ matrices avoiding certain `forbidden' matrices is a well-studied problem in extremal combinatorics with a rich literature and many applications (see a survey by Tardos~\cite{Tardos_2019}).
	In particular, Marcus and Tardos~\cite{MT04} proved:
	
\begin{thm}{\cite{MT04}}\label{thm:MT}
	Let $P$ be a $t \times t$ $0$-$1$ \emph{permutation matrix}, that is, a matrix with exactly one $1$-entry in every row and every column.
	If $M$ is an $n \times n$ $0$-$1$ matrix that avoids $P$, then $M$ contains $O_t(n)$ $1$-entries.
\end{thm}


	Finally, we recall the following result of Fox and Pach~\cite{FP14} which was also mentioned in the Introduction and states that string graphs are degree-bounded.
	
	\begin{thm}[\cite{FP14}]\label{thm:Fox-Pach}
		There is a constant $c$ such that for any positive integers $t$ and $n$, every $K_{t,t}$-free string graph with $n$ vertices has at most $t(\log t)^cn$ edges.
	\end{thm}

	\section{Bounds for $K_{t,t}$-free curve pseudo-visibility graphs}
	\label{sec:pseudo-visibility}
	
	In this section we prove the bounds stated in Theorem~\ref{thm:pseudo-visibility-lb}, Theorem~\ref{thm:pseudo-visibility-ub},
	and Corollary~\ref{cor:K_2,2}.
	
	\subsection{An upper bound for $K_{t,t}$-free curve pseudo-visibility graphs}
	\label{subsec:pseudo-visibility-ub}
	
	In this section we prove Theorem~\ref{thm:pseudo-visibility-ub} stating that every $n$-vertex $K_{t,t}$-free curve pseudo-visibility graph has $O_t(n \log n)$ edges.
	
	\medskip
	Let $M$ be a 0-1 matrix.
	We say that row $i$ and column $j$ \emph{cross} if $M_{ij}=1$ or $M_{ij_1}=M_{ij_2}=M_{i_1j}=M_{i_2j}=1$ for some $i_1 < i < i_2$ and $j_1 < j < j_2$.
	
	\begin{thm}\label{thm:0-1-double-cherry}
		Let $t>0$ be an integer and let $M$ be an $n \times n$ 0-1 matrix without $t$ rows $r_1,r_2,\ldots,r_t$ and $t$ columns $c_1,c_2,\ldots,c_t$
		such that for every $1 \le i,j \le t$ row $r_i$ and column $c_j$ cross.
		Then $M$ has $O_t(n)$ 1-entries.
	\end{thm}
	
	\begin{proof}
		Suppose that $M$ is drawn as an $n \times n$ squares board in which there is a point at the center of every cell that corresponds to a 1-entry.
		For every row of the board draw a horizontal straight-line segment connecting the leftmost and rightmost points in that row (that might be the same point, in which case we get a degenerate segment).
		Similarly, draw for every column a vertical straight-line segment connecting the topmost and bottom-most points in that column.
		Clearly, every point that corresponds to a 1-entry is the intersection point of a horizontal segment and a vertical segment. 
		Therefore, it is enough to bound the number of such intersections, that is, the number of edges in the intersection graph of the at most $2n$ horizontal and vertical segments.
		Since an intersection point at cell $(i,j)$ implies that row $i$ and column $j$ cross, it follows that the intersection graph of the segments is $K_{t,t}$-free.
		By Theorem~\ref{thm:Fox-Pach} this graph has at most $t(\log t)^{O(1)}n$ edges and the claim follows. 
	\end{proof}
		
	Let $G=(V,E,<)$ be an (cyclically-) ordered graph.
	We say that two vertices $u$ and $v$ form a \emph{double cherry} if $(u,v) \in E$ or there are edges $(u,v_1),(u,v_2),(v,u_1),(v,u_2) \in E$ such that $u_1 < u < u_2 < v_1 < v < v_2$.
	Note that in the latter case a double cherry in a cyclically-ordered graph is a special case of crossing sequences from $u$ to $v$ and from $v$ to $u$ (see Figure~\ref{fig:crosing-sequence}).
	\begin{cor}\label{cor:separated}
		Let $t>0$ be an integer and let $G=(A\cup B,E,<)$ be a bipartite ordered graph such that $A<B$.
		If $G$ does not contain two $t$-subsets $A' \subseteq A$ and $B' \subseteq B$, such that every two vertices from different subsets form a double cherry, then $|E|=O_t(|V|)$.
	\end{cor}
	
	\begin{proof}
		We may assume that $|A|=|B|$ by adding isolated vertices if needed.
		Consider the adjacency matrix of $G$ and apply Theorem~\ref{thm:0-1-double-cherry}.
		Note that a cross in this matrix corresponds to a double cherry in $G$ since $A<B$.
	\end{proof}
	
	\begin{cor}\label{cor:non-separated}
		Let $t>0$ be an integer and let $G=(V,E,<)$ be an ordered graph.
		If $G$ does not contain two $t$-subsets of vertices, such that every two vertices from different subsets form a double cherry, then $|E|=O_t(|V|\log|V|)$.
	\end{cor}
	
	\begin{proof}
		Let $c_t$ be the constant hiding in the $O_t(\cdot)$ notation in Corollary~\ref{cor:separated}
		and denote by $f_t(n)$ the maximum number of edges in an $n$-vertex ordered graph without two $t$-subsets of vertices, such that every two vertices from different subsets form a double cherry.
		The claim can be easily proved by induction on $n$, where for the induction step we split the vertices into two (almost) equal subsets $V_1 < V_2$ and get that $f_t(n) \le f_t(\lfloor n/2 \rfloor) + f_t(\lceil n/2 \rceil) + c_t n$, which implies the desired upper bound.
	\end{proof}
	
	\begin{cor}\label{cor:cylically-ordered}
		Let $t>0$ be an integer and let $G=(V,E,<)$ be a cyclically-ordered graph.
		Suppose that $G$ does not contain two $t$-subsets of vertices, such that every two vertices from different subsets form a double cherry. 
		Then $|E|=O_t(|V|\log|V|)$.
		If $G$ is bipartite with bipartition $V=A\dot{\cup}B$ such that $A < B$, then $|E|=O_t(|V|)$.
	\end{cor}
	
	\begin{proof}
		Fix an arbitrary vertex of $G$ as the smallest and consider the resulting ordered graph.
		If it contains two $t$-subsets of vertices, such that every two vertices from different subsets form a double cherry, then so does $G$.
		It follows from Corollary~\ref{cor:non-separated} that $|E|=O_t(|V|\log|V|)$.
		If $G$ is bipartite with bipartition $V=A \dot{\cup} B$ such that $A < B$, then we can fix a vertex of $A$ such that $A<B$ in the resulting linear order and apply Corollary~\ref{cor:separated}.
	\end{proof}

	It follows from Lemma~\ref{lem:crossing-sequences} that if $u$ and $v$ form a double cherry in an ordered curve pseudo-visibility graph, then they are adjacent.
	Therefore, we can now conclude the upper bound for curve pseudo-visibility graphs from Corollary~\ref{cor:cylically-ordered}.
		
	\begin{cor}\label{cor:curve-pseudo}
		Let $t>0$ be an integer and let $G_{\cal L}(K,V)$ be a $K_{t,t}$-free curve pseudo-visibility graph of a set of $n$ points $V$ that lie on a Jordan curve $K$ such that every pair of points in $V$ also lie on a pseudoline from a set of pseudolines $\L$.
		Then $G_{\cal L}(K,V)$ has $O_t(n\log n)$ edges.
		Furthermore, for every $A,B \subseteq V$ such that $A < B$, $G$ has $O_t(|A|+|B|)$ edges $(a,b)$ such that $a\in A$ and $b \in B$. 		
	\end{cor}
		
	\subsection{A lower bound for $K_{t,t}$-free curve pseudo-visibility graphs}
	\label{subsec:pseudo-visibility-lb}		

	In this section we prove Theorem~\ref{thm:pseudo-visibility-lb} stating that there are $K_{3,3}$-free curve pseudo-visibility graphs with $n$ vertices and $\Omega(n\alpha(n))$ edges.
	Then we conclude Corollary~\ref{cor:K_2,2} by which there are also $K_{2,2}$-free curves pseudo-visibility graph with super-linearly many edges.

\begin{figure}
	\centering
	\includegraphics[width= 6cm]{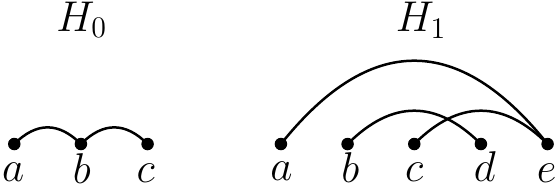}
	\caption{The ordered graphs $H_0$ and $H_1$.}
	\label{fig:F0-F1}
\end{figure}

	\medskip	
	Denote by $H_0$ the ordered graph consisting of three vertices $a < b < c$ and two edges $(a,b)$ and $(b,c)$. 
	Let $H_1$ denote the ordered graph consisting of five vertices $a < b < c < d < e$ and three edges $(a,e)$, $(b,d)$ and $(c,e)$. See Figure~\ref{fig:F0-F1}.

\begin{lem}\label{lem:K_3,3}
	If $G$ is an ordered graph that does not contain neither $H_0$ nor $H_1$ as ordered subgraphs, then $G$ is $K_{3,3}$-free.
\end{lem}

\begin{proof}
	Suppose for contradiction that $G$ contains a copy of $K_{3,3}$ with a bipartition $A=\{a_1,a_2,a_3\}$ and $B=\{b_1,b_2,b_3\}$.
	Since $G$ avoids $H_0$, we have either $A < B$ or $B < A$. 
	Assume without loss of generality that $a_1 < a_2 < a_3 < b_1 < b_2 < b_3$.
	However, then the subgraph $(\{a_1,a_2,a_3,b_2,b_3\},\{(a_1,b_3),(a_3,b_3),(a_2,b_2)\})$ is a copy of $H_1$, a contradiction.
\end{proof}

\begin{thm}[Walczak~\cite{Walczak}]\label{thm:Walczak}
	For every integer $n$ there is an ordered graph with $n$ vertices and $\Omega(n\alpha(n))$ edges which does not contain $H_1$ as an ordered subgraph.
\end{thm}

\begin{proof}
	The lower bound is obtained using the theory of Davenport-Schinzel sequences and refining a construction due to Klazar~\cite{Klazar04}.
	Let $\lambda_s(n)$ denote the maximum length of a \emph{Davenport-Schinzel sequence of order $s$}, that is, a sequence of letters over an alphabet of size $n$ with no two consecutive identical letters and without an alternating sub-sequence of two distinct letters $a,b,a,b,a,b,\ldots$ of length $s+2$.
	By a celebrated result of Hart and Sharir~\cite{HartSharir} $\lambda_3(n)=\Theta(n\alpha(n))$.
	
	Therefore, there is an absolute constant $c_0>0$ such that given $n$ there is a Davenport-Schinzel sequence of length at least $c_0n\alpha(n)$ over the alphabet $[n]$ without a sub-sequence of the form $a,b,a,b,a$.
	Let $v$ be such a sequence and assume without loss of generality that every $i \in [n]$ appears at least twice in $v$ (for otherwise we may add it without forming the forbidden sub-sequence).
	For every $i \in [n]$ let $i^l$ and $i^r$ denote the leftmost and rightmost appearances of $i$ in $v$, respectively.
	Partition $v$ into $2n$ intervals such that the last element in each interval is a leftmost or a rightmost appearance of some letter (and it is the only extreme appearance in the interval).
	Note that if a letter $b$ appears twice within an interval and $a$ appears in between the $b$'s, then this yields the forbidden sub-sequence $a,b,a,b,a$ in $v$. 
	Therefore, the letters within every interval are distinct.
	
	Let $G$ be the ordered graph whose vertices are $\{ i^l, i^r \mid i \in [n]\}$ and their order corresponds to their order in $v$.
	The edges of $G$ are all the pairs $(i^l,j^x)$ such that $i \ne j$ and $i$ appears in the interval which is terminated by $j^x$ ($x=r$ or $x=l$).
	Thus, $G$ has $2n$ vertices and $|v|-2n=\Omega(n\alpha(n))$ edges, where $|v|$ denotes the length of $v$.
	
	We claim that $G$ avoids $H_1$ as an ordered subgraph. 
	Indeed, suppose for contradiction that $G$ contains $H_1$ as a subgraph on the vertices $a^l < b^l < c < d < e$ (note that by construction the first two vertices must correspond to leftmost letters, whereas each of the other vertices may correspond to either leftmost or rightmost letters).
	Thus, $a$ appears in the intervals terminated by $c$ and $e$ and $b$ appears in the interval terminated by $d$.
	However, along with $a^l$ and $b^l$ this implies that $v$ contains the sub-sequence $a,b,a,b,a$ which is a contradiction.
\end{proof}

Walczak~\cite{Walczak} has also shown that the bound in Theorem~\ref{thm:Walczak} is tight.
We are now ready to prove Theorem~\ref{thm:pseudo-visibility-lb}.

\begin{proofof}{Theorem~\ref{thm:pseudo-visibility-lb}}
	Let $n$ be a positive integer.
	We will describe a construction of a Jordan curve $K$ containing a set of $n$ points $V$ and a set of pseudolines $\L$ such that the curve pseudo-visibility graph $G_\L(K,V)$ is $K_{3,3}$-free and has $\Omega(n\alpha(n))$ edges.
	
	By Theorem~\ref{thm:Walczak} there is an $n$-vertex ordered graph $G$ with $\Omega(n\alpha(n))$ edges which avoids $H_1$.
	Remove every edge between two consecutive vertices in $G$. Since there are at most $n-1$ such edges, the number of edges remains $\Omega(n\alpha(n))$.
	Let $G'=(L\cup R,E')$ be the ordered bipartite subgraph of $G'$ guaranteed by Lemma~\ref{lem:left-right-vertices}.
	Observe that $G'$ has $\Omega(n\alpha(n))$ edges and it avoids $H_1$ and $H_0$. 
	Therefore, by Lemma~\ref{lem:K_3,3}, $G'$ is $K_{3,3}$-free. Next we will show that $G'$ can be realized as a curve pseudo-visibility graph.
	
	\paragraph{Drawing $G'$:} Let $P$ be an $x$-monotone polygonal curve we get by connecting $n$ distinct points lying on the lower half of some circle. 
	The points of $V$ correspond to these points such that the $i$th point according to the $x$-order corresponds to the $i$th vertex according to the order of the vertices of $G'$, for every $i=1,2,\ldots,n$. 
	Next, for each edge of $G'$ we connect the corresponding points with a straight-line segment. Henceforth, by a slight abuse of notation, we identify the vertices of $G'$ with the corresponding points and the edges of $G'$ with the corresponding line-segments.
	
	\paragraph{Drawing the non-edges of $G'$:}
	For each non-edge of $G'$ there should be a visibility curve connecting its endpoints which is not entirely inside $K$, since it should not be an edge in the visibility graph.
	Define a partial order on the set of non-edges such that $(a,b) \preceq (c,d)$ if $c \le a < b \le d$.
	It is easy to see that $\preceq$ is indeed a partial order.
	We consider some linear extension of this partial order and draw the non-edges one by one according to this linear extension, that is, $(a,b)$ is drawn before $(c,d)$ if $(a,b) \preceq (c,d)$ (and they are distinct non-edges).
	Let $\varepsilon$ be a very small positive constant and
	suppose that $(a,b)$ is the $i$th non-edge to be drawn, such that $a<b$.
	Denote by $E_{a,b}$ the subset of edges of $G'$ that connect vertices $c$ and $d$ such that $a \le c<d \le b$. Let $P_{a,b}$ be the part of $P$ connecting $a$ and $b$. The visibility curve $\gamma_{a,b}$ between $a$ and $b$ is initially drawn  above and very close --- at distance $\varepsilon i$ --- to the upper envelope of $P_{a,b}\cup E_{a,b}$.
	Note that $\gamma_{a,b}$ does not connect $a$ and $b$ at this point.
	In order to fix that, after drawing all non-edge visibility curves we redraw them at their endpoints such that they now connect the corresponding vertices of $G'$. This can be done in a straightforward manner without introducing crossings between curves that are incident to the same vertex. See Figure~\ref{fig:non-edges} for an illustration.
	\begin{figure}
		\centering
		\subfloat[Before connecting the curves at a vertex]{\includegraphics[width= 5cm]{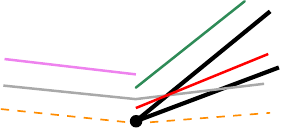}}
		\hspace{15mm}
		\subfloat[Connecting the curves at a vertex, without introducing new crossings between them.]{\includegraphics[width= 5cm]{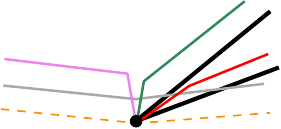}}
		\hspace{5mm}
		\subfloat[Several non-edge visibility curves.]{\includegraphics[width= 12cm]{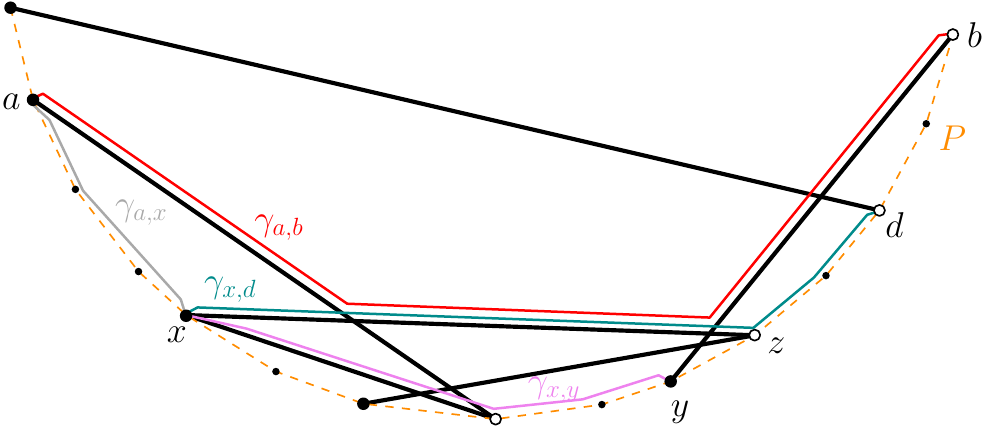}}
		\caption{Drawing the non-edge visibility curves.}
		\label{fig:non-edges}
	\end{figure}
	By a slight abuse of notation we identify henceforth the visibility curves described here with the non-edges of $G'$.
	
	\begin{obs}\label{obs:rotation-system}
		At every vertex $v$ the cyclic order of the edges and non-edges incident to $v$ is the same as their cyclic order would be if the non-edges were drawn as straight-line segments.
	\end{obs}
	
	
	\paragraph{Drawing the curve $K$:} Let $F_{\rm{out}}$ be the outer face of the arrangement of the segments that correspond to the edges of $G'$.
	The boundary of $F_{\rm{out}}$ consists of cycles $C_1,C_2,\ldots,C_l$.
	For each cycle $C_i$ we draw a curve $K_i$ very close to $C_i$ --- at distance at most $\varepsilon/2$ --- such that $K_i$ lies in the interior of $F_{\rm{out}}$ except for the points in $V$ at which $K_i$ touches $C_i$.
	Note that $K_i$ can be (and is) drawn as a polygon such that each of its vertices is either a point in $V$ or a point which is very close to a crossing point of two edges of $G'$.
	If $l>1$ then we need to merge $K_1,\ldots,K_l$ into a single curve, however, we prefer to postpone it for later.
	Figure~\ref{fig:merge-curves} illustrates an example of the curves $K_i$ and hints how they are going to be merged. 
	Notice that $P \setminus V$ lies outside of $\bigcup K_i$ since we assumed that $G'$ has no edge between consecutive vertices.
	
	\medskip
	By construction each drawn edge of $G'$ lies inside $\bigcup K_i$.
	To complete the proof we need show that: (1)~every drawn non-edge crosses $\bigcup K_i$; (2)~the drawn edges and non-edges form a pseudo-segment arrangement; and (3)~this arrangement can be extended into a pseudoline arrangement.
	We prove these facts in the following propositions.
	
	\begin{prop}\label{prop:non-edges-cross-K}		
		Every non-edge visibility curve crosses $\bigcup K_i$.
	\end{prop}
	
	\begin{proof}
		Let $(a,b)$ be a non-edge such that $a < b$.
		Since $\bigcup K_i$ goes arbitrarily close to the boundary of $F_{\rm out}$, it is enough to prove that $\gamma_{a,b}$ intersects $F_{\rm out}$. For this, it is also enough to show that there is a part of $\gamma_{a,b}$ that goes along an edge of $P$, since $P$ lies outside $\bigcup K_i \setminus V$. 
		
		Suppose first that $a$ is a `right' vertex, that is, all its neighbors in $G'$ are to its left. Then the part of $\gamma_{a,b}$ incident to $a$ goes along the edge of $P$ between $a$ and its following vertex along $P$ and we are done, see Figure~\ref{fig:a-right}.
		\begin{figure}
			\centering
			\includegraphics[width=10cm]{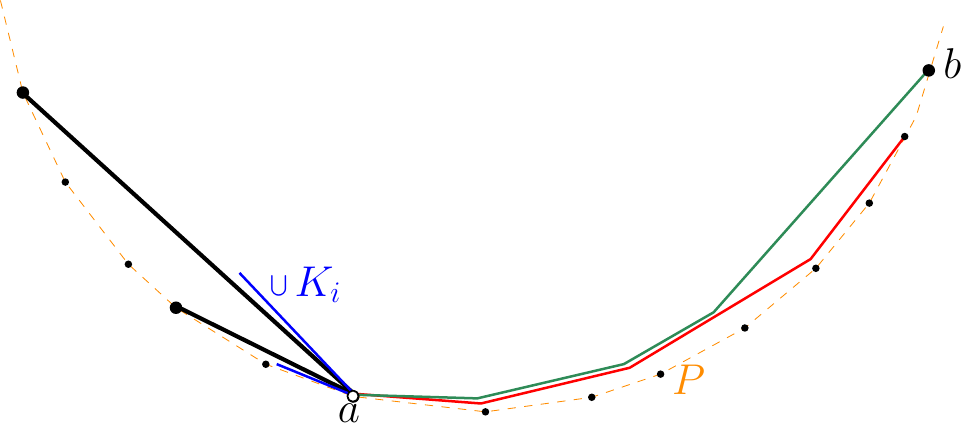}
			\caption{If $(a,b)$ is a non-edge, $a<b$ and $a$ is a `right' vertex, then $\gamma_{a,b}$ runs along the edge of $P$ between $a$ and its following vertex along $P$. Thus, $\gamma_{a,b}$ is outside $\bigcup K_i$ locally near $a$.}
			\label{fig:a-right}
		\end{figure}
		If $b$ is a `left' vertex, that is, all its neighbors in $G'$ are to its right, then we conclude by a symmetric argument.
		
		Suppose now that $a$ is a left-vertex, $b$ is a right-vertex and the upper envelope of $P_{a,b}\cup E_{a,b}$, denote it by $U_{a,b}$, does not run along any edge of $P_{a,b}$, that is, it is actually the upper envelope of $E_{a,b}$, see Figure~\ref{fig:non-edges}(c) for an example.
		Observe that there is no point $v\in V\setminus\{a,b\}$ close to $U_{a,b}$, since this would imply that there are two edges incident to $v$ which form a copy of $H_0$ in $G'$. 
		Therefore, we may assume that $U_{a,b}$ is never close to a point $v\in V\setminus\{a,b\}$.
		Consider the leftmost vertex $y$ such that $a \le y < b$ and $(y,b) \in E_{a,b}$. As $(a,b)$ is a non-edge, we have that $a<y$.
		Such a vertex must exist by our assumptions. 
		Notice that $\gamma_{a,b}$ goes closely above $(y,b)$ in the vicinity of $b$. 
		Also, since $y$ is not close to $U_{a,b}$, there must be at least one further edge $(x,z)$ with $a \le x < y < z < b$. 
		Note that there cannot be an edge $(v,b)$ in $G'$ with $v<y$.
		Indeed, if such an edge exists, then necessarily $v<a$ for otherwise we would choose $v$ instead of $y$. However, if $v<a$ then the edges $(v,b)$, $(y,b)$ and $(x,z)$ form a copy of $H_1$ in $G'$, which is impossible. 
		Finally, since $b$ is a right-vertex, there are no edges $(b,v)$ such that $b<v$ and therefore we conclude that the boundary of $F_{\rm out}$ must go along $(y,b)$ in the vicinity of $b$ and thus $\gamma_{a,b}$ must intersect $\bigcup K_i$ there.  
		
		We note that in the last case, i.e., when $U_{a,b}$ never runs close to $P$, $\gamma_{ab}$  could actually be drawn as a straight-line segment connecting $a$ and $b$.
	\end{proof}
	
	\begin{prop}\label{prop:arrangement}
		The edges and non-edge curves form a pseudo-segment arrangement.
		Furthermore, this arrangement can be extended into a pseudoline arrangement.
	\end{prop}

	\begin{proof}
		Clearly every two drawn edges intersect at most once since they are drawn as straight-line segments. 
		Let $\gamma_{a,b}$ (where $a<b$) be a non-edge curve that intersects an edge $e = (x,y)$ (where $x<y$).
		Since all edges and non-edges are drawn as $x$-monotone curves, the interval of vertices from $a$ to $b$ and the interval of vertices from $x$ to $y$ are not disjoint. Consider first the case that the two curves share an endpoint.
		Assume without loss of generality that $x=a$.
		If $y>b$, then the curves do not intersect at another point but $a$,
		because the interior of $\gamma_{a,b}$ lies below the straight-line segment from $a$ to $b$ whereas the interior of $e$ lies above it in this case.
		Otherwise if $y<b$, then $e \in E_{a,b}$. Therefore, the curves do not intersect at another point but $a$, since $\gamma_{a,b}$ lies above the upper envelope of $E_{a,b}$.
		
		Therefore, we may assume that the two curves do not share an endpoint.
		Since $\gamma_{a,b}$ is a convex curve apart from near its endpoints, it follows that if it intersects $e$ more than once, then it must intersect $e$ exactly twice and that $a<x<y<b$.
		However, in this case $e \in E_{a,b}$ and we can conclude that $e$ does not intersect $\gamma_{a,b}$, since $\gamma_{a,b}$ lies above the upper envelope of $E_{a,b}$.
		
		It remains to consider the case of two non-edge curves $\gamma_{a,b}$ and $\gamma_{c,d}$.
		We assume without loss of generality that $a < b$, $c < d$ and $a \le c$.
		If $a<b\le c<d$, then the two curves are trivially disjoint apart from possibly one common endpoint.
		If $a\le c<d\le b$, then $(c,d) \preceq (a,b)$ and $E_{c,d} \subseteq E_{a,b}$ which implies that $\gamma_{a,b}$ is drawn later than $\gamma_{c,d}$ and above it except for possibly at a common endpoint. 
		
		Finally, suppose that $a<c<b<d$ and $\gamma_{a,b}$ and $\gamma_{c,d}$ intersect more than once. Note that these intersection points are crossing points, because the curves do not share endpoints.
		Then, there has to be a crossing point $q$ at which $\gamma_{c,d}$ gets below $\gamma_{a,b}$ when we follow the curves from left to right.
		Since $c$ lies below $\gamma_{a,b}$, going rightwards from $c$ the curve $\gamma_{c,d}$ may only get above $\gamma_{a,b}$.
		Therefore, $q$ is not in the vicinity of $c$.
		Similarly, because $b$ lies below $\gamma_{c,d}$, going rightwards towards $b$ the curve $\gamma_{a,b}$ may only get below $\gamma_{c,d}$.
		Therefore, $q$ is not in the vicinity of $b$ either.
		Thus, $q$ must be in the vicinity of a crossing point $q'$ of two edges $e \in E_{a,b}$ and $f \in E_{c,d}$, such that at $q'$ the edge $f$ gets below the edge $e$ (going from left to right). By definition of $\gamma_{a,b}$ and $\gamma_{c,d}$, in order for these curves to actually go along the respective edges $e$ and $f$, we must also have $e \in E_{a,b} \setminus E_{c,d}$ and $f \in E_{c,d} \setminus E_{a,b}$.
		Because $e \in E_{a,b} \setminus E_{c,d}$ and its right endpoint is between $q'$ and $b$ (including $b$), its left endpoint must be to the left of $c$ (as it is not in $E_{c,d}$).
		However, the left endpoint of $f$ is to the right of $c$ or at $c$ since $f \in E_{c,d}$, which implies that $e$ and $f$ cross once more to the left of $q'$ (see Figure~\ref{fig:e-f}). 
		That is of course impossible.
		\begin{figure}
			\centering
			\includegraphics[width=10cm]{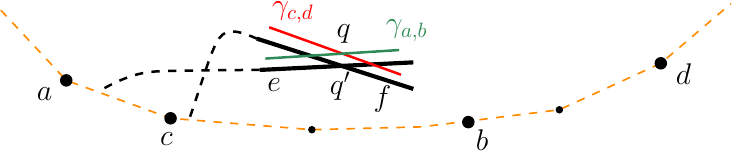}
			\caption{If $\gamma_{a,b}$ and $\gamma_{c,d}$ intersect more than once and $a < c < b < d$, then there is $e \in E_{a,b} \setminus E_{c,d}$ and $f \in E_{c,d} \setminus E_{a,b}$ such that $f$ goes below $e$ which implies that $e$ and $f$ cross once more.}
			\label{fig:e-f}
		\end{figure}
		
		\medskip
		We have concluded that the drawn edges and non-edges form a pseudo-segment arrangement.
		From Observation~\ref{obs:rotation-system} it follows that we can extend this arrangement into a pseudoline arrangement as follows.
		For every edge and non-edge $(a,b)$ add $\ell_{a,b}\setminus \overline{ab}$ to the curve connecting $a$ and $b$, where $\ell_{a,b}$ is the line through $a$ and $b$ and $\overline{ab}$ is the straight-line segment connecting $a$ and $b$.
	\end{proof}
	
	Finally, recall that $\bigcup K_i$ may consist of several curves which should be merged into a single curve $K$.
	This can be done as follows. 
	First, draw a horizontal straight-line segment $h$ below $P$ such that the projection of $h$ on the $x$-axis contains the projection of $P$ on the $x$-axis.
	Then, for every $K_i$ pick an arbitrary left-vertex $v \in V$ on the boundary of $K_i$ and a point $p$ on $K_i$ very close to $v$ such that the interior of $K_i$ is locally to the left of $\overrightarrow{vp}$. 	Then it is possible to draw a vertical straight-line segment from $p$ to $h$ such that this segment does not intersect $\bigcup K_i$ at any point but $p$.
	We do that for every $K_i$.
	Then we slightly `inflate' all these vertical segments and $h$ such that we obtain a single curve $K$, see Figure~\ref{fig:merge-curves} for an illustration.
	\begin{figure}
		\centering
		\includegraphics[width=12cm]{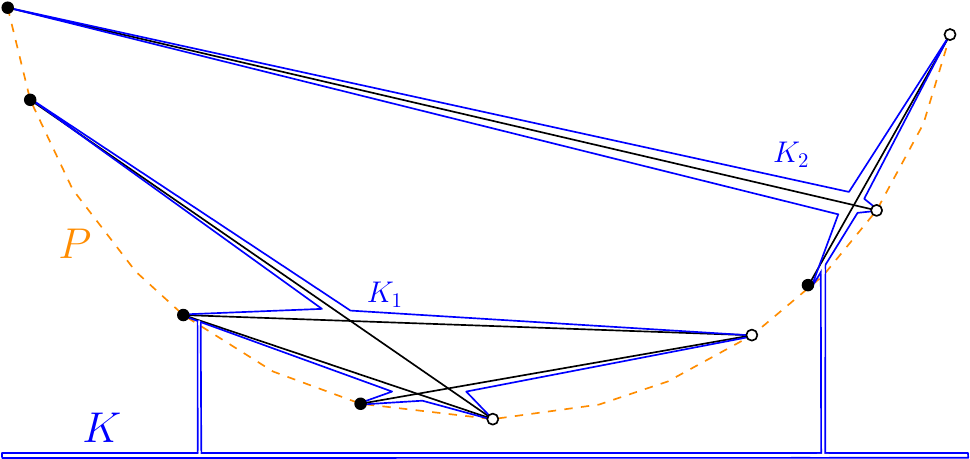}
		\caption{Merging $K_1, K_2, \ldots, K_l$ into a single curve $K$.}
		\label{fig:merge-curves}
	\end{figure}
	Clearly, every drawn edge of $G'$ lies in the interior of $K$ and every non-edge curve crosses the boundary of $K$.
	Let $\L$ be the pseudoline arrangement guaranteed by Proposition~\ref{prop:arrangement}.
	Then $G_{\L}(K,V)$ is the required curve pseudo-visibility graph.
\end{proofof}
	
	We conclude this section with a proof of Corollary~\ref{cor:K_2,2} which states that there is no constant $c_2$ such that every $n$-vertex $K_{2,2}$-free curve pseudo-visibility graph has at most $c_2n$ edges.
	This follows from a result of McCarty~\cite{McCarty21} which implies that in order to show that a hereditary family of graphs is degree-bounded, it is enough to consider its $K_{2,2}$-free members.
	For completeness, these arguments follow.
	
	\begin{thm}[{\cite[Theorem 1]{McCarty21}}]\label{thm:McCarty}
	There is a function $f_1:\mathbb{N}\rightarrow\mathbb{N}$ so that for every $t\in\mathbb{N}$, every bipartite graph with average degree at least $f_1(t)$ has either a $K_{t,t}$-subgraph or an induced subgraph of average degree at least $t$ and girth at least $6$.		
	\end{thm}
	
	\begin{proofof}{Corollary~\ref{cor:K_2,2}}
	Suppose for contradiction that there is a constant $c_2>0$ such that every $n$-vertex $K_{2,2}$-free curve pseudo-visibility graph has at most $c_2n$ edges. Set $t=2\lceil c_2+1\rceil$ and take a bipartite $K_{3,3}$-free curve pseudo-visibility graph with average degree at least $f_1(t)$ (there is such a graph by Theorem~\ref{thm:pseudo-visibility-lb}). 
	This graph is $K_{t,t}$-free since it is $K_{3,3}$-free and $t\ge 3$, therefore it follows from Theorem~\ref{thm:McCarty} that it has an induced subgraph of average degree at least $t$ and girth at least $6$, denote it by $G'=(V',E')$. 
	Note that $G'$ is also a curve pseudo-visibility graph because pseudo-visibility graphs are hereditary. Since the girth of $G'$ is at least $6$, it is $K_{2,2}$-free. 
	However, its average degree is at least $t$, thus $|E'| \ge t|V'|/2 = \lceil c_2+1 \rceil |V'| > c_2|V'|$, a contradiction. 
	\end{proofof}

	\section{Bounds for $K_{t,t}$-free polygon visibility graphs}
	\label{sec:polygon}

	In this section we prove the upper bound on the size of $K_{t,t}$-free polygon visibility graphs and the better bounds for the visibility graph of star-shaped and $x$-monotone polygons.
	
	\subsection{$K_{t,t}$-free polygon visibility graphs}
	\label{subsec:polygon}

	Recall that in a polygon visibility graph every vertex of the given polygon is a vertex of its visibility graph. Using this fact we can show that as soon as a polygon visibility graph contains crossing edges, it contains $K_4$ (and hence $K_{2,2}$) as a subgraph.

\begin{thm}\label{thm:K_4}
	Let $P$ be an $n$-gon and let $G$ be the visibility graph of $P$.
	If $G$ has more than $2n-3$ edges, then $G$ contains $K_{4}$ as a subgraph.
\end{thm}

\begin{proof}
	Suppose that the size of $G$ is greater than $2n-3$.
	Then $G$ has crossing edges, since an $n$-vertex outerplanar graph has at most $2n-3$ edges.
	Let $(a,b)$ and $(c,d)$ be a pair of crossing edges such that the area of the (convex) quadrilateral determined by their endpoints is minimal.
	Suppose without loss of generality that $a <_P c <_P b <_P d$ and denote by $x$ the crossing point of the two edges.
	
	If the sides of the quadrilateral $acbd$ are edges of $G$, then they form a copy of $K_{4}$.
	Therefore, suppose that at least one of these sides is crossed by $P$. 
	It follows that at least one of the four triangles determined by $x$ and a side of $acbd$ contains a vertex of $P$.
	Assume without loss of generality that $\triangle acx$ is such a triangle. 
	Among the vertices of $P$ that are inside $\triangle acx$ let $y$ be a vertex such that $\angle yax$ is minimal.
	If we extend $\overline{ay}$ beyond $y$ until it hits $(c,d)$ at a point $y'$, then by the definition of $y$ the interior of $\triangle ay'x$ does not contain a vertex of $P$ (refer to Figure~\ref{fig:K_2,2}). 
	\begin{figure}
		\centering
		\includegraphics[width= 6cm]{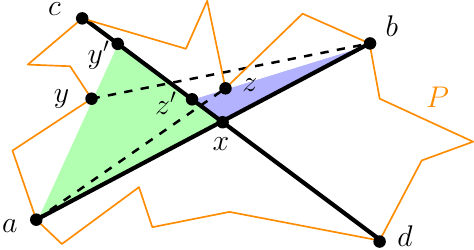}
		\caption{A polygon visibility graph with crossing edges must contain a copy of $K_{4}$.}
		\label{fig:K_2,2}
	\end{figure}
	
	The segment $\overline{yb}$ crosses $(c,d)$ and their endpoints determine a quadrilateral of smaller area than $acbd$.
	It follows that $\triangle ayb \cap \triangle cxb$ must contain a vertex of $P$.
	Among these vertices let $z$ be a vertex such that $\angle zbx$ is minimal.
	If we extend $\overline{bz}$ beyond $z$ until it hits $(c,d)$ at a point $z'$, then by the definition of $z$ the interior of $\triangle bz'x$ does not contain a vertex of $P$.
	This implies that $\overline{az}$ forms an edge of $G$ since it is contained in $\triangle ay'x \cup \triangle bz'x$.
	However, $(a,z)$ crosses $(c,d)$ and their endpoints determine a quadrilateral of smaller area than $acbd$, a contradiction. 		
\end{proof}

	Next we prove the bound on the size of a $K_{t,t}$-free polygon visibility graph which is stated in Theorem~\ref{thm:polygon-ub}. For this we need the following.

\begin{prop}\label{prop:bowtie2}
	Let $P$ be a simple polygon and let $r_1,r_2,c_1,c_2,c_3,c_4$ be vertices of $P$ that appear in this clockwise order on the boundary of $P$.
	If $r_1$ sees $c_1$ and $c_3$ and $r_2$ sees $c_2$ and $c_4$, then $P$ has a vertex $v$ such that $c_2 \le_P v \le_P c_3$ and $v$ sees $r_1$ and $r_2$.
\end{prop}

\begin{proof}
	Considering the four visibility edges $(r_1,c_1)$,  $(r_1,c_3)$, $(r_2,c_2)$ and $(r_2,c_4)$,  
	it follows from the cyclic order of the vertices that $(r_1,c_1)$ crosses $(r_2,c_2)$ and $(r_2,c_4)$ and that $(r_1,c_3)$ crosses only $(r_2,c_4)$ (since topologically these segments are diagonals of the topological disk $P$).
	It also follows that $\overrightarrow{r_1c_1}$ hits $(r_2,c_4)$ at a point $x$ and then $(r_2,c_2)$ at a point $y$ and that  $\overrightarrow{r_2c_4}$ hits $(r_1,c_1)$ at $x$ and then $(r_1,c_3)$ at a point $z$, see Figure~\ref{fig:bowtie}.
	\begin{figure}[h]
	\centering
	\includegraphics[width= 8cm]{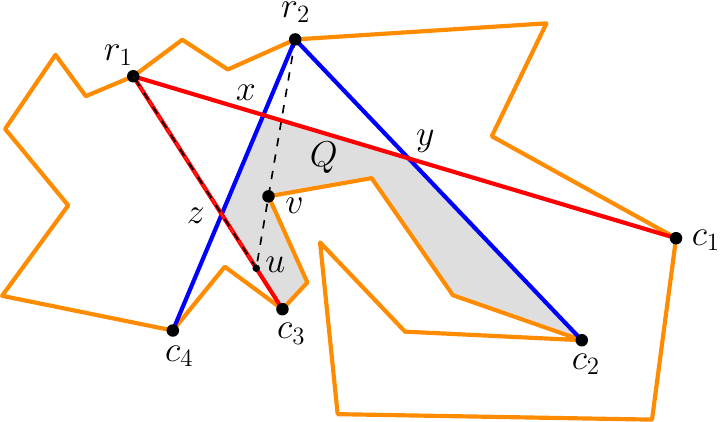}
	\caption{An illustration for the proof of Proposition~\ref{prop:bowtie2}.}
	\label{fig:bowtie}
\end{figure}
	Let $Q$ be the region bounded by $c_2 \leadsto_P c_3$ and the edge-segments $\overline{c_3z}$, $\overline{zx}$, $\overline{xy}$ and $\overline{yc_2}$.
	Clearly, $Q$ contains no vertex of $P$ in its interior.
	Fix a point $u$ at $z$ and observe that $u$ sees $r_1$ and $r_2$ and continues to see them as we move it continuously along $\overline{c_3z}$ until either $u$ and $c_3$ coincide or $\overline{ur_2}$ intersects $c_2 \leadsto_P c_3$ at a vertex of $P$.
	This is the sought vertex $v$ in the latter case whereas $v=c_3$ in the former case.
\end{proof}

\begin{cor}\label{cor:bowtie}
	Let $P$ be a simple polygon and let $r_1,r_2,\ldots,r_k,c_1,c_2,\ldots,c_{2k}$ be vertices of $P$ that appear in this clockwise order on the boundary of $P$.
	If $r_i$ sees $c_i$ and $c_{i+k}$ for every $i=1,2,\ldots,k$, then $P$ has a vertex $v$ such that $c_{k} \le_P v \le_P c_{k+1}$ and $v$ sees $r_j$ for every $j=1,2,\ldots,k$.
\end{cor}

\begin{proof}
	By applying Proposition~\ref{prop:bowtie2} for $r_1,r_k,c_1,c_k,c_{k+1}$ and $c_{2k}$ we conclude that there is a vertex $v$ that sees $r_1$ and $r_k$ such that $c_{k} \le_P v \le_P c_{k+1}$.
	Consider $r_j$ for some $1 < j < k$.
	It follows from the cyclic order of the vertices that the edges $(r_1,v)$ and $(r_j,c_{k+j})$ cross and so do the edges $(r_k,v)$ and $(r_j,c_j)$.
	Therefore, $v$ and $r_j$ form a double cherry and thus by Lemma~\ref{lem:crossing-sequences} they are adjacent in the visibility graph of $P$. 
\end{proof}

\begin{proofof}{Theorem~\ref{thm:polygon-ub}}
	Let $G=(V,E)$ be a $K_{t,t}$-free visibility graph of a simple $n$-gon $P$ and
	let $v_1,v_2,\ldots,v_n$ be the vertices of $P$ listed in their clockwise order starting from an arbitrary vertex $v_1$. 
	Denote by $G'=(L \cup R, E')$ the ordered bipartite subgraph guaranteed by Lemma~\ref{lem:left-right-vertices} and let $A'$ be its adjacency matrix. 
	That is, the rows of $A'$ correspond to vertices in $L$ according to their order and its columns correspond to vertices in $R$ according to their order.
	
	Denote by $M_t$ the $t \times (4t-2)$ 0-1 matrix we get by: taking the identity matrix $I_t$; concatenating to it $t-1$ copies of the $t \times 2$ matrix with exactly two 1-entries, at cells $(1,1)$ and $(t,2)$; and concatenating to the result the identity matrix $I_t$.
	For example, 
	
	$$ M_4 = \begin{pmatrix}
		1 &   &   &   & 1 &   & 1 &   & 1 &   & 1 \\
		& 1 &   &   &   &   &   &   &   &   &   & 1 \\
		&   & 1 &   &   &   &   &   &   &   &   &   & 1 \\
		&   &   & 1 &   & 1 &   & 1 &   & 1 &   &   &   & 1 
	\end{pmatrix}  .$$

	Given an $m \times n$ $0$-$1$ matrix $M$, we denote by $M^+$ the matrix we get by adding to $M$ a new last row and a new first column, setting the $(m+1,1)$ entry to $1$ and all the other new entries to $0$.
		
	Consider $M^+_t$ and observe that it has exactly one $1$-entry at each column. It follows from a result of Klazar~\cite{Kla92} that an $n \times n$ $0$-$1$ matrix avoiding such a matrix has $O(n2^{\alpha(n)^{O(1)}})$ $1$-entries.\footnote{This is mentioned in \cite{klazarthesis,tardos01}, for a full proof see~\cite[Theorem 2.15]{keszeghmsc}.} Therefore, the number of $1$-entries in an $n \times n$ $0$-$1$ matrix which avoids $M^+_t$ is $O(n2^{\alpha(n)^{c_t}})$ for some constant $c_t$.
	Thus, it is enough to show that $A'$ avoids $M^+_t$.
	
	Suppose for contradiction that $A'$ contains $M^+_t$ as a submatrix, let 
	$r_{1} < r_{2} < \cdots < r_{t+1}$ be the vertices of $G'$ that correspond to the rows of this submatrix and let $l_{0} < l_{1} < \cdots < l_{4t-2}$ be those vertices that correspond to its columns.
	Since $(r_{t+1},l_{0}) \in E'$, it follows that $r_{t+1} < l_{0}$.
	Thus, we have a subgraph of $G'$ whose vertices are $r_{1} < r_{2} < \cdots < r_{t} < l_{1} < l_{2} < \cdots < l_{4t-2}$ and whose edges correspond to $M_t$, see Figure~\ref{fig:Klazar} for an illustration.
	\begin{figure}
		\centering
		\includegraphics[width= 10cm]{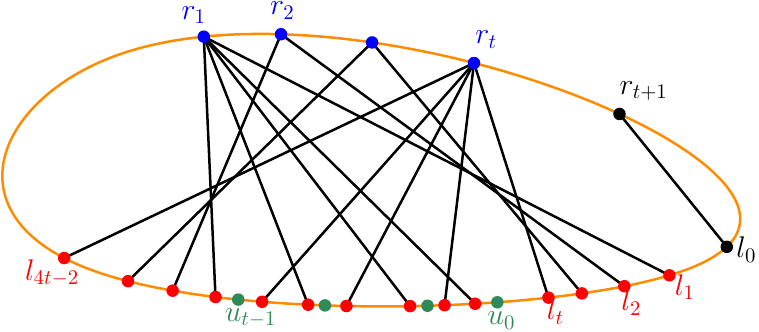}
		\caption{An illustration for the proof of Theorem~\ref{thm:polygon-ub}: $A'$ contains $M^+_t$ as a submatrix.}
		\label{fig:Klazar}		
	\end{figure}
	Now for every $s=0,2,\ldots,t-1$, by applying Corollary~\ref{cor:bowtie} on $r_1, r_{2},\ldots, r_{t}$, $l_{1}, l_{2},\ldots, l_{t-1}$, $l_{t+2s}$, $l_{t+2s+1}$, $l_{3t}, \ldots, l_{4t-2}$ we conclude that there is a vertex $u_s$ such that $l_{t+2s} \le u_s \le l_{t+2s+1}$ and $u_s$ is a neighbor of each of $r_{1}, r_{2}, \ldots, r_{t}$.
	Therefore, $\{r_{1},\ldots,r_{t}\} \cup \{u_0,\ldots,u_{t-1}\}$ is a bipartition of a $K_{t,t}$ subgraph in $G$, a contradiction.
\end{proofof}

	\subsection{A linear upper bound for star-shaped polygons}
		
	\begin{proof}[Proof of Theorem~\ref{thm:star-shaped}]
		Let $P$ be a star-shaped polygon and let $V$ be a set of $n$ points on its boundary.
		Denote by $G=(V,E)$ the visibility graph of $V$ with respect to $P$.
		Let $p$ be a point in $P$ which sees every point in $P$ and let $\ell$ be a non-vertical line through $p$ containing no point of $V$, except possibly for $p$ if $p \in V$.
		Denote by $V_1$ and $V_2$ the points in $V$ above and below $\ell$, respectively.
		It follows from Corollary~\ref{cor:curve-pseudo} that $G$ has $O_t(n)$ edges $(v_1,v_2)$ such that $v_1 \in V_1$ and $v_2 \in V_2$.
		If $p \in V$, then there are exactly $n-1$ edges such that $p$ is one of their endpoints.
		It remains to count edges that do not intersect $\ell$.
		
		Let $G_1=(V_1,E_1)$ be the subgraph induced by $V_1$ and let $v_1,v_2,\ldots,v_{n_1}$ be its vertices listed as they appear in clockwise order (starting from $\ell$).
		We now bound the number of edges in $G_1$, the analysis of the subgraph induced by $V_2$ is analogous.
		\begin{prop}\label{prop:star-shaped}
			If $(v_a,v_b), (v_c,v_d) \in E_1$ are crossing edges such that $a<c<b<d$, then $(v_a,v_d) \in E_1$.
		\end{prop}
		
		\begin{proof}
			Let $q$ be the crossing point of $(v_a,v_b)$ and $(v_c,v_d)$.
			The point $p$ sees $v_a, v_c,v_b, v_d$ in this angular order, since these points appear in that  order on the boundary of $P$ and $p$ sees every point on the boundary of $P$.
			Therefore, $v_b$ and $v_c$ lie in the cone bounded by the rays $\overrightarrow{pv_a}$ and $\overrightarrow{pv_d}$.
			If $v_b \in \triangle pv_av_d$, then $v_c \in \triangle pv_av_d$ as well since $(v_a,v_b)$ and $(v_c,v_d)$ cross (see Figure~\ref{fig:star-shaped-triangle}).
		\begin{figure}
		\centering
		\subfloat[$v_b$ and $v_c$ cannot lie in $\triangle pv_av_d$.]{\includegraphics[width= 6.5cm]{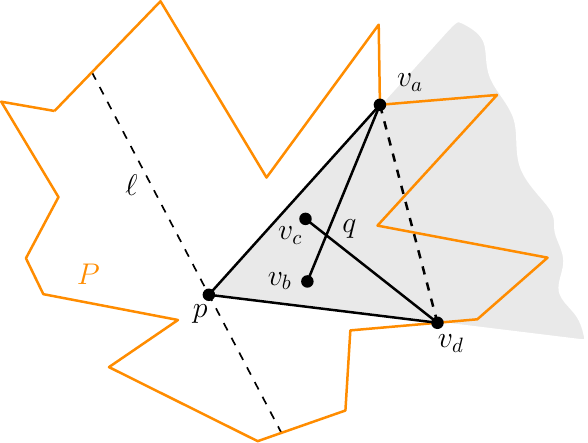}\label{fig:star-shaped-triangle}}
		\hspace{ 5mm}
		\subfloat[$(v_a,v_d)$ is an edge of the visibility graph]{\includegraphics[width= 6cm]{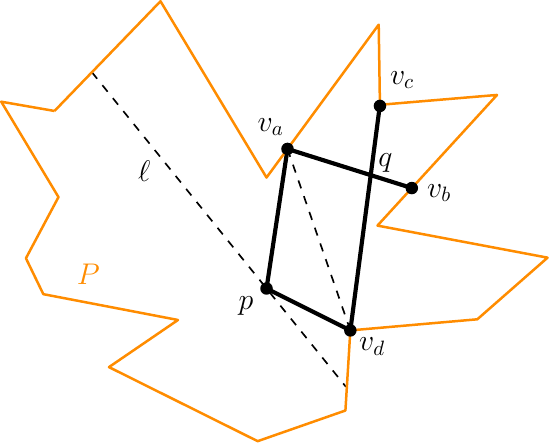}\label{fig:star-shaped-v_av_d}}
		\hspace{5mm}
		\caption{An illustration for the proof of Proposition~\ref{prop:star-shaped}: $v_a,v_c,v_b,v_d$ appear in this order on the boundary of the star-shaped polygon $P$. $(v_a,v_b)$ and $(v_c,v_d)$ are crossing edges.}
		\label{fig:star-shaped}
	\end{figure}
			However, then the quadrilateral $pv_aqv_d$ contains in its interior vertices of $P$ (namely, $v_b$ and $v_c$), which is impossible since the sides of this quadrilateral are visibility edges or segments of such edges.
			
			It follows that $p$ and $\{v_b,v_c\}$ lie on different sides of the line through $v_a$ and $v_d$.
			Therefore, $p$ and $q$ also lie on different sides of that line.
			Consequently, $\overline{v_av_d}$ lies in the quadrilateral $pv_aqv_d$ and it follows that $(v_a,v_d) \in E_1$ (see Figure~\ref{fig:star-shaped-v_av_d}).
		\end{proof}
		
		We claim that $G_1$ has no $2t$ pairwise crossing edges and therefore by Theorem~\ref{thm:Capoyleas-Pach} we have $|E_1| = O_t(n_1)$.
		Indeed, let $(a_1,b_1), (a_2,b_2), \ldots, (a_{2t},b_{2t})$ be pairwise crossing edges, such that their clockwise order (starting from $\ell$) is $a_1,a_2,\ldots,a_{2t},b_1,b_2,\ldots,b_{2t}$.
		Then it follows from Proposition~\ref{prop:star-shaped} that $\{a_1,a_2,\ldots,a_t\} \cup \{b_{t+1},b_{t+2},\ldots,b_{2t}\}$ is a bipartition of a $K_{t,t}$ subgraph.
	\end{proof}
	
	\subsection{A linear upper bound for $x$-monotone polygons}
		
	\begin{proof}[Proof of Theorem~\ref{thm:x-monotone}]
		Let $P$ be an $x$-monotone polygon and let $V$ be a set of $n$ points on its boundary.
		Let $G=(V,E)$ be the visibility graph of $V$ with respect to $P$.
		To simplify the presentation we assume that no two vertices of $P$ share the same $x$-coordinate.
		Denote by $P_u$ the \emph{upper chain} of $P$, that is, the clockwise polygonal chain from its leftmost vertex to its rightmost vertex.
		Similarly, denote by $P_l$ be the \emph{lower chain} of $P$, that is, the counterclockwise polygonal chain from its leftmost vertex to its rightmost vertex.
		Set $V_l = V \cap P_l$ and $V_u = V \cap P_u$.
		By Corollary~\ref{cor:curve-pseudo} there are $O_t(n)$ edges in $G$ such that one of their endpoints is in $V_l$ and their other endpoint is in $V_u$.
		It remains to consider the edges in the subgraphs induced by $V_l$ and by $V_u$.
		
		Let $G^l=(V_l,E_l)$ be the ordered subgraph of $G$ induced by $V_l$ and let $v_1<v_2<\ldots<v_k$ be its vertices ordered as they appear along $P_l$ from left to right (where $k = |V_l|$).
		Let $G'_l=(L'\cup R', E'_l)$ be the ordered bipartite subgraph of $G_l$ guaranteed by Lemma~\ref{lem:left-right-vertices} and let $A'_l$ be its adjacency matrix. That is,
		the rows of $A'_l$ correspond to vertices in $L'$ according to their order and its rows correspond to vertices in $R'$ according to their order.
		It is enough to show that $A'_l$ has $O_t(n)$ $1$-entries.
		Set $M_t = \left((I_{2t}^+)^\intercal\right)^+$.\footnote{Recall that for an $m \times n$ $0$-$1$ matrix $M$, we denote by $M^+$ the matrix we get by adding to $M$ a new last row and a new first column, setting the $(m+1,1)$ entry to $1$ and all the other new entries to $0$.}
			 For example, 
		\begin{equation*}
			M_2=
			\begin{pmatrix}
				& & & & & 1 \\
				& 1 & & & & \\
				& & 1 & & & \\
				& & & 1 & & \\
				& & & & 1 & \\
				1 & & & & &
			\end{pmatrix}.
		\end{equation*}
		We claim that $A'_l$ does not contain $M_t$ as a submatrix.
		Suppose for contradiction that it does, let 
		$l_{0} < l_{1} < \cdots < l_{2t+1}$ be the vertices of $L'$ that correspond to the rows of this submatrix and let $r_{0} < r_{1} < \cdots r_{2t+1}$ be the vertices of $R'$ that correspond to its columns.
		Since $(l_{2t+1},r_{0}) \in E'_l$, it follows that $l_{2t+1} < r_{0}$.
		Thus, we have a subgraph of $G_l$ whose vertices are $l_{0} < l_{1} < \cdots < l_{2t+1} < r_{0} < r_{1} < \cdots < r_{2t+1}$ and whose edges are $(l_{0},r_{2t+1})$, $(l_{2t+1},r_{0})$, $(l_{1},r_{1}),(l_{2},r_{2}),\ldots,(l_{2t},r_{2t})$, see Figure~\ref{fig:matching} for an illustration.
		\begin{figure}[!h]
			\centering
			\includegraphics[width=12cm]{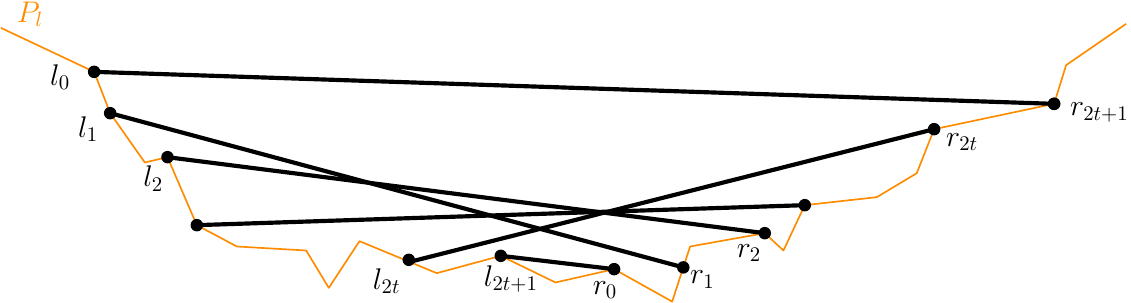}
			\caption{An illustration for the proof of Theorem~\ref{thm:x-monotone}: $A'_l$ contains $M_t$ as a submatrix.}
			\label{fig:matching}
		\end{figure}
		Observe that the edges $(l_{1},r_{1}),(l_{2},r_{2}),\ldots,(l_{2t},r_{2t+1})$ are pairwise crossing.
		Consider two such edges  $(l_{i},r_{i})$ and $(l_{j},r_{j})$, $i<j$, and the straight-line segment $\overline{l_{i}r_{j}}$.
		The part of $P_l$ from $l_{i}$ to $r_{j}$ lies below the lower envelope of these two edges and therefore cannot cross $\overline{l_{i}r_{j}}$.
		It follows from the $x$-monotonicity of $P$ and the fact that $\overline{l_{i}r_{j}}$ lies below the edge $(l_{0},r_{2t+1})$ that $\overline{l_{i}r_{j}}$ is not crossed by $P_u$ or other parts of $P_l$. 
		Therefore,  $(l_{i},r_{j}) $ is an edge in $G$.
		Hence, $\{l_{1},l_{2},\ldots,l_{t}\} \cup \{r_{t+1},r_{t+2},\ldots,r_{2t}\}$ is a bipartition of a $K_{t,t}$ subgraph of $G$, a contradiction.
		
		Therefore $A'_l$ does not contain the permutation matrix $M_t$ as a submatrix and it follows from a result of Theorem~\ref{thm:MT} that there are $O_t(k)$ 1-entries in $A'_l$. 
		Hence, $G$ contains $O_t(|V_l|)$ edges between vertices in $V_l$.
		Similarly, there are $O_t(|V_u|)$ edges between vertices in $V_u$ and the theorem follows.
	\end{proof}

	\paragraph{Remarks.} We keep thinking of $t$ as fixed and use the $O_t(\cdot)$ notation for our asymptotic bounds. Suppose that we analyze the constant $C_t$ hiding in the $O_t(\cdot)$ notation in the proof of Theorem~\ref{thm:x-monotone} as a function of $t$.
	Then the usage of the Marcus-Tardos Theorem (Theorem~\ref{thm:MT}) implies the non-polynomial bound $C_t = O\left(t^4{{t^2}\choose{t}}\right)$.
	However, we have proved that visibility graphs of points on the boundary of an $x$-monotone polygon are degree-bounded, and these graphs are clearly hereditary. Therefore, by the result of Gir\~ao and Hunter~\cite{GH24} that was mentioned in the Introduction it follows that $C_t$ can be bounded from above by a polynomial function of $t$.
	Moreover, Hunter, Milojevi\'c, Sudakov and Tomon~\cite{Hunter25} recently improved the upper bounds in Theorems~\ref{thm:star-shaped} and~\ref{thm:x-monotone} to $O(tn)$.

	\paragraph{Acknowledgments.} We are grateful to Bartosz Walczak for his permission to include his proof of Theorem~\ref{thm:Walczak} in this paper. We also thank G\'abor Tardos for informing us of this result.


	\footnotesize
	\bibliographystyle{plainurl}
	\bibliography{polygon-visibility}

@article{CP92,
	title = {A {T}urán-type theorem on chords of a convex polygon},
	journal = {Journal of Combinatorial Theory, Series B},
	volume = {56},
	number = {1},
	pages = {9--15},
	year = {1992},
	issn = {0095-8956},
	doi = {https://doi.org/10.1016/0095-8956(92)90003-G},
	url = {https://www.sciencedirect.com/science/article/pii/009589569290003G},
	author = {Vasilis Capoyleas and János Pach},
}

@article{Coloring23,
	title = {Coloring polygon visibility graphs and their generalizations},
	journal = {Journal of Combinatorial Theory, Series B},
	volume = {161},
	pages = {268--300},
	year = {2023},
	issn = {0095-8956},
	doi = {https://doi.org/10.1016/j.jctb.2023.02.010},
	url = {https://www.sciencedirect.com/science/article/pii/S0095895623000187},
	author = {James Davies and Tomasz Krawczyk and Rose McCarty and Bartosz Walczak},
}

@article{FP14, 
	title={Applications of a New Separator Theorem for String Graphs}, 	
	volume={23}, 
	DOI={10.1017/S0963548313000412}, 
	number={1}, 
	journal={Combinatorics, Probability and Computing}, 
	author={Fox, Jacob and Pach, J\'anos}, 
	year={2014}, 
	pages={66--74}}

@article{MT04,
	title = {Excluded permutation matrices and the {S}tanley–{W}ilf conjecture},
	journal = {Journal of Combinatorial Theory, Series A},
	volume = {107},
	number = {1},
	pages = {153--160},
	year = {2004},
	issn = {0097-3165},
	doi = {https://doi.org/10.1016/j.jcta.2004.04.002},
	url = {https://www.sciencedirect.com/science/article/pii/S0097316504000512},
	author = {Adam Marcus and Gábor Tardos},
}

@article{Kla92,
	title = {A general upper bound in extremal theory of sequences},
	journal = {Comment.\ Math.\ Univ.\ Carolin.},
	volume = {33},
	number = {4},
	pages = {737--746},
	year = {1992},
	author = {Martin Klazar},
}

@misc{Walczak,
	author = "Bartosz Walczak",
	howpublished = "personal communication",
}

@Article{HartSharir,
	author={Hart, Sergiu
	and Sharir, Micha},
	title={Nonlinearity of {D}avenport-{S}chinzel sequences and of generalized path compression schemes},
	journal={Combinatorica},
	year={1986},
	month={Jun},
	day={01},
	volume={6},
	number={2},
	pages={151--177},
	issn={1439-6912},
	doi={10.1007/BF02579170},
	url={https://doi.org/10.1007/BF02579170}
}

@article{Klazar04,
	title = {Extremal problems for ordered (hyper)graphs: applications of {D}avenport–{S}chinzel sequences},
	journal = {European Journal of Combinatorics},
	volume = {25},
	number = {1},
	pages = {125--140},
	year = {2004},
	issn = {0195-6698},
	doi = {https://doi.org/10.1016/j.ejc.2003.05.001},
	url = {https://www.sciencedirect.com/science/article/pii/S0195669803001239},
	author = {Martin Klazar}
}

@article{degree-boundeness,
	title = {A survey of degree-boundedness},
	journal = {European Journal of Combinatorics},
	year = {2024},
	issn = {0195-6698},
	doi = {https://doi.org/10.1016/j.ejc.2024.104092},
	url = {https://www.sciencedirect.com/science/article/pii/S019566982400177X},
	author = {Xiying Du and Rose McCarty}
}

@article{JanzerP24,
	author       = {Oliver Janzer and Cosmin Pohoata},
	title        = {On the {Z}arankiewicz Problem for Graphs with Bounded {VC}-Dimension},
	journal      = {Comb.},
	volume       = {44},
	number       = {4},
	pages        = {839--848},
	year         = {2024},
	url          = {https://doi.org/10.1007/s00493-024-00095-2},
	doi          = {10.1007/S00493-024-00095-2},
	bibsource    = {dblp computer science bibliography, https://dblp.org}
}

@article{FranklK21,
	author       = {N{\'{o}}ra Frankl and Andrey Kupavskii},
	title        = {On the {E}rd{\H{o}}s-{P}urdy problem and the {Z}arankiewitz problem for
	semialgebraic graphs},
	journal      = {CoRR},
	volume       = {abs/2112.10245},
	year         = {2021},
	url          = {https://arxiv.org/abs/2112.10245},
	eprinttype    = {arXiv},
	eprint       = {2112.10245},
	bibsource    = {dblp computer science bibliography, https://dblp.org}
}

@article{FoxPSSZ,
	author       = {Jacob Fox and J\'anos Pach and Adam Sheffer and Andrew Suk and Joshua Zahl},
	title        = {A semi-algebraic version of {Z}arankiewicz's problem},
	journal      = {J.\ Eur.\ Math.\ Soc.},
	volume       = {19},
	number       = {6},
	pages        = {1785--1810},
	year         = {2017},
	url          = {https://ems.press/journals/jems/articles/14758},
}

@misc{GH24,
	title={Induced subdivisions in ${K}_{s,s}$-free graphs with polynomial average degree}, 
	author={António Girão and Zach Hunter},
	year={2024},
	eprint={2310.18452},
	archivePrefix={arXiv},
	primaryClass={math.CO},
	url={https://arxiv.org/abs/2310.18452}, 
}

@article{BBCD24,
	author       = {Romain Bourneuf and Matija Buci\'{c} and Linda Cook and James Davies},
	title        = {On Polynomial Degree-Boundedness},
	journal      = {Advances in Combinatorics},
	year         = {2024},
	url          = {https://doi.org/10.19086/aic.2024.5},
}

@article{HunterMST25,
	author       = {Zach Hunter and
	Aleksa Milojevic and
	Benny Sudakov and
	Istv{\'{a}}n Tomon},
	title        = {K{\H{o}}v{\'{a}}ri-{S}{\'{o}}s-{T}ur{\'{a}}n theorem for hereditary families},
	journal      = {J. Comb. Theory {B}},
	volume       = {172},
	pages        = {168--197},
	year         = {2025},
	url          = {https://doi.org/10.1016/j.jctb.2024.12.009},
	doi          = {10.1016/J.JCTB.2024.12.009},
	bibsource    = {dblp computer science bibliography, https://dblp.org}
}

@article{Kvri1954OnAP,
	title={On a problem of {K.} {Z}arankiewicz},
	author={Tam{\'a}s Kőv{\'a}ri and Vera T. S{\'o}s and Paul Tur{\'a}n},
	journal={Colloquium Mathematicum},
	year={1954},
	volume={3},
	pages={50-57},
	url={https://api.semanticscholar.org/CorpusID:118684929}
}

@inproceedings{KellerS2024,
	title={Zarankiewicz's Problem via $\epsilon$-$t$-Nets},
	author={Chaya Keller and Shakhar Smorodinsky},
	booktitle={International Symposium on Computational Geometry},
	year={2024},
	url={https://api.semanticscholar.org/CorpusID:270286589}
}

@article{Szemerdi1983ExtremalPI,
	title={Extremal problems in discrete geometry},
	author={Endre Szemer{\'e}di and William T. Trotter},
	journal={Combinatorica},
	year={1983},
	volume={3},
	pages={381-392},
	url={https://api.semanticscholar.org/CorpusID:1750834}
}

@misc{smorodinsky2024,
	title={A survey of {Z}arankiewicz problem in geometry}, 
	author={Shakhar Smorodinsky},
	year={2024},
	eprint={2410.03702},
	archivePrefix={arXiv},
	primaryClass={math.HO},
	url={https://arxiv.org/abs/2410.03702}, 
}

@article{CAGIRICI2024,
	title = {On colourability of polygon visibility graphs},
	journal = {European Journal of Combinatorics},
	volume = {117},
	pages = {103820},
	year = {2024},
	note = {Selected papers of EuroComb 2019},
	issn = {0195-6698},
	doi = {https://doi.org/10.1016/j.ejc.2023.103820},
	url = {https://www.sciencedirect.com/science/article/pii/S0195669823001373},
	author = {Onur Çağırıcı and Petr Hliněný and Bodhayan Roy}
}

@article{GHOSH2007,
	title = {Computing the maximum clique in the visibility graph of a simple polygon},
	journal = {Journal of Discrete Algorithms},
	volume = {5},
	number = {3},
	pages = {524-532},
	year = {2007},
	note = {Selected papers from Ad Hoc Now 2005},
	issn = {1570-8667},
	doi = {https://doi.org/10.1016/j.jda.2006.09.004},
	url = {https://www.sciencedirect.com/science/article/pii/S1570866706000955},
	author = {Subir Kumar Ghosh and Thomas Caton Shermer and Binay Kumar Bhattacharya and Partha Pratim Goswami}
}

@article{Ghosh97,
	author       = {Subir Kumar Ghosh},
	title        = {On Recognizing and Characterizing Visibility Graphs of Simple Polygons},
	journal      = {Discret. Comput. Geom.},
	volume       = {17},
	number       = {2},
	pages        = {143--162},
	year         = {1997},
	url          = {https://doi.org/10.1007/BF02770871},
	doi          = {10.1007/BF02770871},
	bibsource    = {dblp computer science bibliography, https://dblp.org}
}

@article{Hershberger89,
	author       = {John Hershberger},
	title        = {An Optimal Visibility Graph Algorithm for Triangulated Simple Polygons},
	journal      = {Algorithmica},
	volume       = {4},
	number       = {1},
	pages        = {141--155},
	year         = {1989},
	url          = {https://doi.org/10.1007/BF01553883},
	doi          = {10.1007/BF01553883},
	bibsource    = {dblp computer science bibliography, https://dblp.org}
}

@article{GhoshG13,
	author       = {Subir Kumar Ghosh and
	Partha P. Goswami},
	title        = {Unsolved problems in visibility graphs of points, segments, and polygons},
	journal      = {{ACM} Comput. Surv.},
	volume       = {46},
	number       = {2},
	pages        = {22:1--22:29},
	year         = {2013},
	url          = {https://doi.org/10.1145/2543581.2543589},
	doi          = {10.1145/2543581.2543589},
	bibsource    = {dblp computer science bibliography, https://dblp.org}
}

@inbook{Tardos_2019, place={Cambridge}, series={London Mathematical Society Lecture Note Series}, title={Extremal theory of vertex or edge ordered graphs}, booktitle={Surveys in Combinatorics 2019}, publisher={Cambridge University Press}, author={Tardos, Gábor}, editor={Lo, Allan and Mycroft, Richard and Perarnau, Guillem and Treglown, AndrewEditors}, year={2019}, pages={221–236}, collection={London Mathematical Society Lecture Note Series}}

@Article{Agarwal1994,
author={Agarwal, P. K.
and Alon, N.
and Aronov, B.
and Suri, S.},
title={Can visibility graphs Be represented compactly?},
journal={Discrete {\&} Computational Geometry},
year={1994},
month={Sep},
day={01},
volume={12},
number={3},
pages={347-365},
issn={1432-0444},
doi={10.1007/BF02574385},
url={https://doi.org/10.1007/BF02574385}
}

@article{keszeghmsc,
  title={Forbidden submatrices in 0-1 matrices},
  author={Keszegh, Bal{\'a}zs},
  journal={Msc Thesis},
  year={2005}
}

@article{tardos01,
title = {On 0–1 matrices and small excluded submatrices},
journal = {Journal of Combinatorial Theory, Series A},
volume = {111},
number = {2},
pages = {266-288},
year = {2005},
issn = {0097-3165},
doi = {https://doi.org/10.1016/j.jcta.2004.11.015},
url = {https://www.sciencedirect.com/science/article/pii/S009731650500004X},
author = {Gábor Tardos}
}

@article{klazarthesis,
  title={Enumerative and extremal combinatorics of a containment relation of
partitions and hypergraphs},
  author={Klazar, Martin},
  journal={Habilitation Thesis},
  year=2001
}

@article{Do2019,
	author = {Do, Thao},
	title = {Representation Complexities of Semialgebraic Graphs},
	journal = {SIAM Journal on Discrete Mathematics},
	volume = {33},
	number = {4},
	pages = {1864--1877},
	year = {2019},
	publisher = {Society for Industrial and Applied Mathematics},
	doi = {10.1137/18M1221606}
}

@article{Hunter25,
	author       = {Zach Hunter and
	Aleksa Milojevi\'c and
	Benny Sudakov and
	Istv{\'{a}}n Tomon},
	title        = {C\({}_{\mbox{4}}\)-free subgraphs of high degree with geometric applications},
	journal      = {CoRR},
	volume       = {abs/2506.23942},
	year         = {2025},
	url          = {https://doi.org/10.48550/arXiv.2506.23942},
	doi          = {10.48550/ARXIV.2506.23942},
	eprinttype    = {arXiv},
	eprint       = {2506.23942},
	bibsource    = {dblp computer science bibliography, https://dblp.org}
}

@article{McCarty21,
	author = {McCarty, Rose},
	title = {Dense Induced Subgraphs of Dense Bipartite Graphs},
	journal = {SIAM Journal on Discrete Mathematics},
	volume = {35},
	number = {2},
	pages = {661--667},
	year = {2021},
	doi = {10.1137/20M1370744},
	URL = { 
	https://doi.org/10.1137/20M1370744
	},
	eprint = { 
	https://doi.org/10.1137/20M1370744	
	}
}
	
\end{document}